\documentclass{emulateapj}
\pdfoutput=1
\usepackage{natbib}
\usepackage{times}
\usepackage{graphicx}
\usepackage{amsmath}

\usepackage[colorlinks=true,citecolor=black,filecolor=blue,linkcolor=blue,urlcolor=blue,pdftex]{hyperref}

\begin{document}

\shorttitle{Once Upon a Time in Sgr A West}
\shortauthors{Kistler}

\title{A Tale of Two Pulsars and the Origin of TeV Gamma Rays from the Galactic Center}

\author{Matthew D. Kistler}
\affiliation{Kavli Institute for Particle Astrophysics and Cosmology, Stanford University, SLAC National Accelerator Laboratory, Menlo Park, CA 94025}


\begin{abstract}
The Galactic Center (GC) has been long known to host gamma-ray emission detected to $>\,$10~TeV.  HESS data now points to two plausible origins: the supermassive black hole (perhaps with $\gtrsim\,$PeV cosmic rays and neutrinos) or high-energy electrons from the putative X-ray pulsar wind nebula G359.95--0.04 observed by {\it Chandra} and {\it NuSTAR}.  We show that if the magnetic field experienced by PWN electrons is near the several mG ambient field strength suggested by radio observations of the nearby GC magnetar SGR J1745-29, synchrotron losses constrain the TeV gamma-ray output to be far below the data.  Accounting for the peculiar geometry of GC infrared emission, we also find that the requisite TeV flux could be reached if the PWN is $\sim\,$1~pc from Sgr~A$^*$ and the magnetic field is two orders of magnitude weaker, a scenario that we discuss in relation to recent data and theoretical developments.  Otherwise, Sgr~A$^*$ is left, which would then be a PeV link to other AGN.

\end{abstract}


\section{Introduction}
The central parsec has naturally attracted much attention, with a long history of observations utilizing the windows available to see through the Galactic disk \citep{Genzel2010}.  Among the many phenomena with varying degrees of bizarreness within this region is the substantial population of compact remnants from a long history of stellar birth and death, i.e., black holes, neutron stars, and white dwarfs.  However, radio surveys for pulsars have come up empty (see \citealt{Dexter2014,Chennamangalam2014}).

Evidence for a fairly powerful pulsar arose from {\it Chandra} observations that resolved a cometary nebula with a non-thermal X-ray spectrum, G359.95--0.04 (hereafter G359), suggestive of a pulsar wind nebula (PWN) shaped by ram pressure at a projected distance of $\sim\,$0.3~pc from Sgr~A$^*$ \citep{Wang2006,Muno2008}.  The spectral steepening away from the head of the nebula is consistent with synchrotron cooling of electrons/positrons, with a 2--10~keV luminosity of $L_{\rm keV}\!\sim\! 10^{34}\,$erg~s$^{-1}$ \citep{Wang2006}; however, thus far no pulsations have been reported in any band.

An intriguing possible connection is with the hard, unresolved TeV gamma-ray emission coincident with the GC seen by air Cherenkov telescopes with $L_{\rm TeV}\!\sim\! 10^{35}\,$erg~s$^{-1}$ \citep{Aharonian2004,Albert2006,Archer2014}.  \citet{Wang2006} suggested that multi-TeV $e^\pm$ producing synchrotron in the {\it Chandra} band are associated with the TeV gamma rays, since the ratio of magnetic field to ambient photon energy density could be low near the GC.  The more detailed model of \citet{Hinton2007} used a field strength of $105\,\mu$G to fit both {\it Chandra} and HESS data.  HESS has since provided the most precise spectrum of this source \citep{Aharonian2009} and with a much improved localization of the centroid of the emission narrowed the focus to two counterparts: Sgr~A$^*$ and G359 \citep{Acero2010}.

If G359 can be demonstrated to {\it not} be a substantial TeV emitter, this would leave Sgr~A$^*$ as the best candidate standing, from which there is no shortage of interesting scenarios for producing TeV gamma rays (see Section~\ref{concl}).  Some of these yield high-energy neutrinos, which may well account for the PeV energy event seen by IceCube from the GC direction \citep{Aartsen2013}.  Moreover, while the Sgr~A$^*$ supermassive black hole is unique in the Milky Way, it is common to massive galaxies in general and would prove an important linkage to the larger high-energy universe \citep{Kistler2015b,Kistler2015c}.

We evaluate the viability of G359 as the source of the GC TeV gamma rays after taking into account a variety of recent data.  The most stunning twist was the (unrelated) soft gamma repeater, SGR J1745-29, discovered very near Sgr~A$^*$ by {\it Swift} \citep{Kennea2013}.  {\it NuSTAR} detected 3.76~s X-ray pulsations \citep{Mori2013} indicating a magnetar at $\sim\!2.4\,$arcsec ($\sim\!0.1\,$pc projected) from Sgr~A$^*$ \citep{Rea2013}.  Of special interest is the pulsed radio emission detected with unusually large dispersion and rotation measures \citep{Shannon2013}.  \citet{Eatough2013} concluded that the Faraday rotation results from the diffuse hot gas around the GC, with a large implied field strength of $B\!\gtrsim\!8$~mG at the 0.1~pc scale.

It is not spoiling too much to say here that we first find that, if a mG magnetic field is germane to the synchrotron emitting $e^\pm$, the TeV flux obtained by normalizing to the measured X-ray flux falls well below the HESS data.  This can be simply understood from the fact that increasing $B$ from $105\,\mu$G as in \citet{Hinton2007} to $>\!1000\,\mu$G increases the synchrotron loss rate by a factor of $>\!100$, which does spoil the achievement of $L_{\rm TeV}/L_{\rm keV}\!\approx\! 10$.

Less obvious is whether conditions for TeV gamma-ray production are necessarily unsatisfactory, keeping in mind that the positional evidence for proximity to Sgr~A$^*$ is more circumstantial than for objects with, e.g., orbital data.  We have considered a wide variety of possibilities associated with G395 and present here several representative scenarios to illustrate a plausible range.

We focus on positions within the central parsec, since at larger distances the benefit of bright, compact infrared emission potentially producing a large $L_{\rm TeV}/L_{\rm keV}$ ratio is lost.  Recent high spatial resolution measurements of the relevant infrared emission have revealed structures within the central parsec.  We account for this anisotropy in TeV gamma-ray production via inverse Compton scattering and destruction due to $\gamma \!-\! \gamma$ pair production with the toy model of the GC photon field from \citet{Kistler2015}.

We examine further implications, including the consistency of the implied PWN properties with recent particle acceleration research.  The {\it NuSTAR} discovery of very hard X-ray emission extending to $>\,$40~keV throughout this region has a peak near G395 \citep{Perez2015,Mori2015}.  We discuss the implications of synchrotron extension extending into this range, suggestive of $\gtrsim\,$100~TeV $e^\pm$, and the connection to gamma rays.  Also, while the GC has long also been seen in GeV gamma rays, with the best observations now being from {\it Fermi} \citep{Acero2015}, it is not clear what is producing this emission.  In the course of addressing requirements on the pulsar wind, we also will estimate the implied properties of pulsed GeV gamma rays from G359.

\section{Emission from a PWN at the Galactic Center: I}
We begin by examining simplified scenarios for producing X-rays and TeV gamma rays from G359.  To do so requires some basic data.  One of these is the pulsar age.  The head of the G359 nebula is 8.7~arcsec from Sgr~A$^*$ ($\sim\! 0.3\,$pc projected) and the tail approaches to with $\sim\!4$~arcsec \citep{Wang2006}.  Examining positions of known GC supernova remnants \citep{Ponti2015} and backtracking based on the direction implied by the PWN orientation reveals no obvious correlation (SNR Sgr~A East is in the opposite direction).

\citet{Wang2006} suggest an association with the IRS~13 stellar complex, which, for a $\sim\,$few hundred km~s$^{-1}$ relative velocity would give an age of $\sim\!10^3\,$yr.  Lacking a characteristic spin-down age or SNR, strictly speaking we only know for certain that it was not born during the period of neutrino monitoring for Galactic SNe beginning in the 1980's (e.g., \citealt{Alexeyev2002,Ikeda2007}).  For instance, no supernova is recorded even from the $\lesssim\! 150\,$yr old GC SNR G1.9+0.3 \citep{Green2008}.  \citet{Hinton2007} assumed a constant $e^\pm$ luminosity over the past $10^4\,$yr.  For concreteness, we here use continuous injection for $\tau \!=\! 10^3\,$yr.

To evolve the spectrum of these $e^\pm$, we use the techniques described in \citet{Kistler2015} along with their central parsec photon field model.  Once we obtain the present spectrum, we will evaluate the spectra of inverse Compton (IC) photons using each angular-dependent background component, integrating from the vantage point of the PWN over angles with respect to the direction pointing at Earth to obtain the IC spectrum as
\begin{equation}
       \frac{dN_i}{dE_\gamma}  = \mathcal{E}_i  \int_{E_\gamma}^{E_{\rm max}} dE_e \frac{dN_e}{dE_0} \int d\Omega\, \frac{dN_{\rm ani}}{dE_\gamma} \ell_i(\theta,\phi)
    \,,
\label{ICspec}
\end{equation}
where $\mathcal{E}_i \!=\! L_i / (4\pi c\, u_{\rm BB,i} V_i)$, with $u_{\rm BB,i}$ the blackbody energy density for a given $T_i$.  We obtain fluxes $\varphi_i(E_\gamma)$ from scattering on each background component taking $d_{\rm GC} \!=\! 8.5\,$kpc and assuming that the $e^\pm$ population is isotropic and relativistic beaming is not important.

We are left to determine the magnetic field relevant for the synchrotron energy loss rate and spectrum.  \citet{Eatough2013} arrived at a value of $B\!\gtrsim\! 8 \,(26\,{\rm cm}^{-3}/n_0)\,$mG, where $n \!=\! n_0 (0.4\,{\rm pc}/r)$ in the hot gas near Sgr~A$^*$, with $n_0 \!=\! 26\,$cm$^{-3}$ obtained from their modeling of the GC plasma.  One could simply assume that $B\!\sim\!8$~mG throughout the entire central parsec.  The implications of such a scenario can be readily extrapolated from the results of considering two possibilities for a declining field with distance from Sgr~A$^*$.  The first has an equal $B^2$ per radial shell, $B_1(r) \!=\! 8 \, (0.12\,{\rm pc}/r)\,$mG.  The second falls more steeply as $B_2(r) \!=\! 8 \, (0.12\,{\rm pc}/r)^2\,$mG.

Even the smallest of the ambient values discussed previously are much larger than typically encountered by ram pressure confined PWNe far out in the Galactic disk.  Is this value typical of what is going on within G359?  We will assume here that the ambient field strength is encountered by the $e^\pm$ after being accelerated with no field compression, discussing first evidence from the X-ray data.

\begin{figure}[t!]
\hspace*{-0.2cm}
\includegraphics[width=1.04\columnwidth,clip=true]{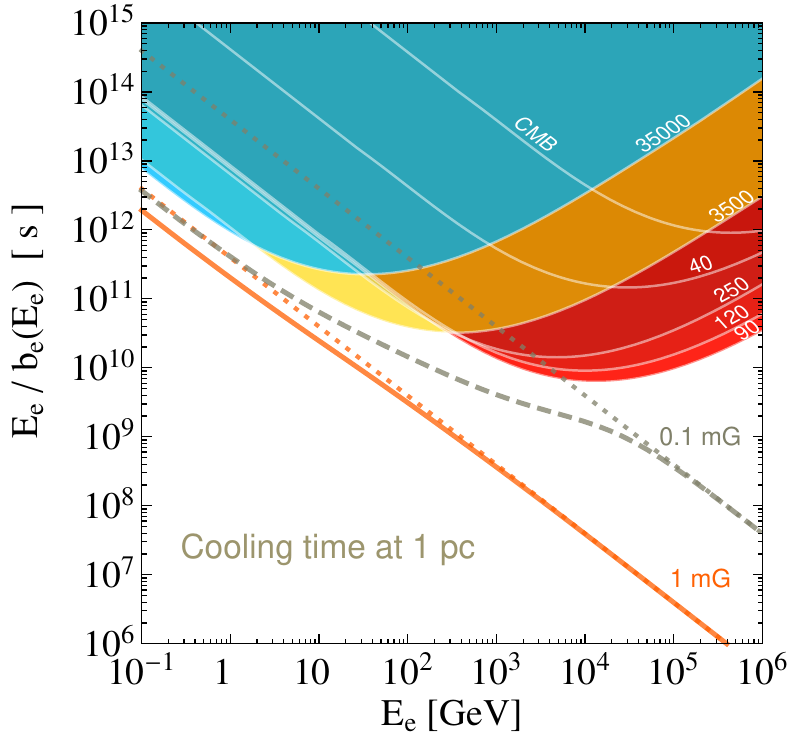}
\caption{Cooling time of electrons at a distance from Sgr~A$^*$ of 1~pc ({\it solid lines}) shown for cases with synchrotron losses for two different field strengths ({\it dashed lines}) and inverse Compton losses on each Galactic Center background component of the photon field from \citet{Kistler2015}.\\
\label{bEfig}}
\end{figure}

\begin{figure*}[t!]
\includegraphics[width=2.09\columnwidth,clip=true]{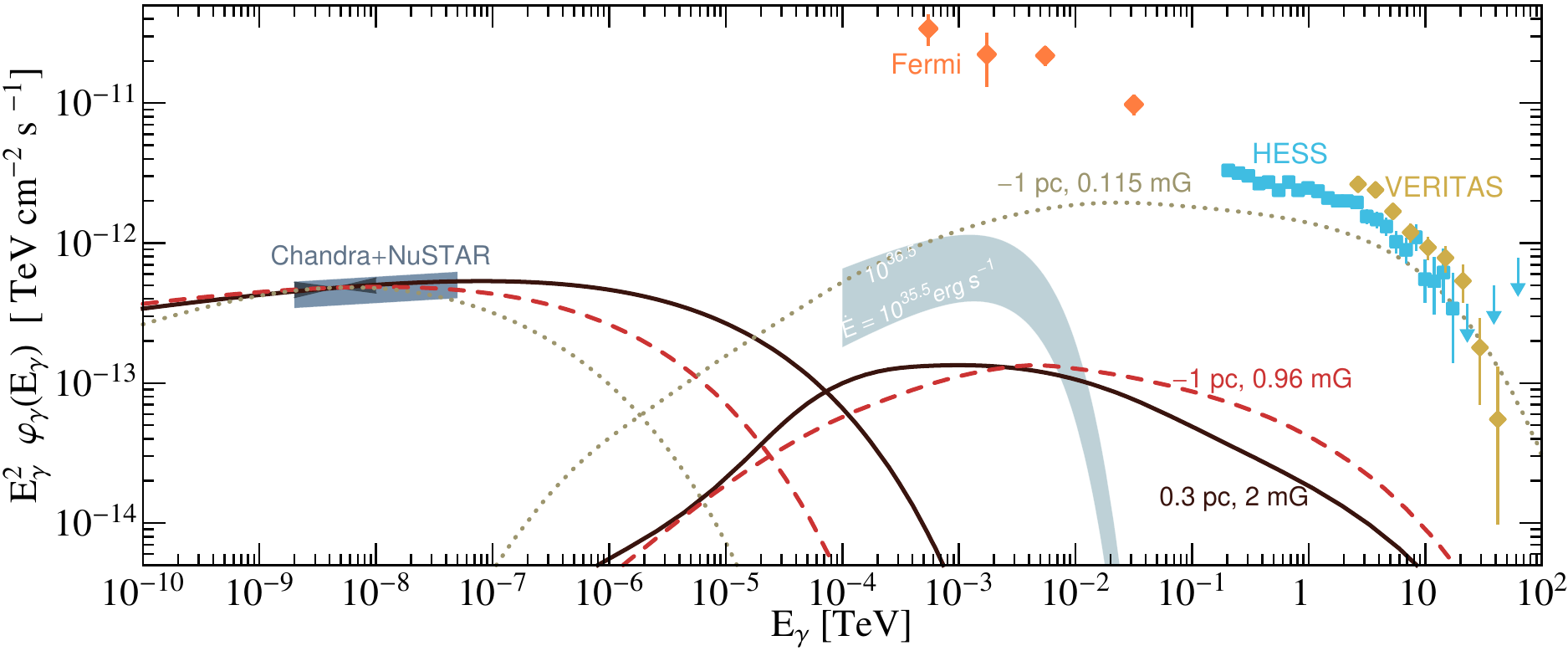}
\caption{Synchrotron and inverse Compton spectra for models using a distance of 0.3~pc from Sgr~A$^*$ with 2~mG field ({\it solid lines}) or 1~pc (behind Sgr~A$^*$) with either 0.96~mG ({\it dashed lines}) or 0.115~mG ({\it dotted lines}) and parameters in Table~\ref{tab:pwn}.  These are compared to {\it Chandra} + {\it NuSTAR} G359 (\citealt{Mori2015}), {\it Chandra} ({\it bowtie}; \citealt{Wang2006}), the Galactic Center HESS \citep{Abramowski2016} and VERITAS \citep{Archer2016} source and {\it Fermi} \citep{Acero2015} GeV source.  We also estimate the pulsed $0.1 \!-\! 100\,$GeV flux from the G359 pulsar based on the inferred PWN spindown $\dot{E} \!\sim\! 3 \!-\! 30 \!\times\! 10^{35}\,$erg~s$^{-1}$ ({\it band}).\\
\label{speca}}
\end{figure*}

In Fig.~\ref{bEfig}, we show both the total cooling rate and as broken into components for the photon field at a distance of 1~pc from Sgr~A$^*$ for two different values of $B$.  These characteristic cooling times show how the importance of IC varies compared to synchrotron and the Klein-Nishina (KN) suppression via the break in the IC curves with increasing $E_e$.

As mentioned earlier, the spectral steepening with increasing distance away from the head of the nebula measured by \citet{Wang2006} is a keystone of the PWN interpretation.  They determined that to achieve a synchrotron cooling time for electrons emitting $\sim\,$keV X-rays comparable to the travel time along the full length of the nebula requires a field of $B \!\sim\! 0.8 \,(v_f \sin{i}/c_s)^{2/3}\,$mG, where $i$ is the inclination of the PWN to the observer, $v_f$ is the flow velocity, and $c_s \!=\! c/\sqrt{3}$ is the sound speed of relativistic plasma.  For the nominal parameters this is consistent with a field strength in the mG range.

We can also consider the Cannonball, a PWN with similar morphology located $\sim\,$20~pc from Sgr~A$^*$ and presumed to be associated with the Sgr~A East SNR.  The Cannonball has been seen in hard X-rays by {\it NuSTAR}, from which a field of $\sim\! 313 \!-\! 530\,\mu$G was derived by \citet{Nynka2013}, consistent with the $\sim\! 0.3\,$mG field estimate from radio equipartition arguments in \citet{Zhao2013}.

\citet{Wang2006} reported a power-law fit to the G359 X-ray spectrum with a slope of $\sim\!-1.4$ near the PWN head steepening to $\sim\! -2$, which motivates a choice of spectral index for the synchrotron emitting $e^\pm$ of $\alpha \!\gtrsim\! -2$.  We take this as a starting point for the source spectrum, described via a smoothly-broken power law with an exponential cutoff
\begin{equation}
      \frac{dN_e}{dE}  =   f_e
       \left[\left(E/E_1\right)^{\alpha \eta} + \left(E/E_1\right)^{\beta \eta} \right]^{1/\eta} e^{-E/E_c}\,,
\label{fit}
\end{equation}
with $\alpha$ and $\beta$ the slopes, a break at $E_1$, cutoff energy $E_c$, and $\eta \!=\! -10$ for a sharp break.  In these scenarios, we assume a break at $E_1 \!=\! 1\,$GeV, below which the slope is $\beta \!=\! 2$.  Knowing only the projected distance, we consider three locations for G359: $d \!=\! 0.3\,$pc; in front of Sgr~A$^*$ at $d \!=\! 1\,$pc; or to the rear of Sgr~A$^*$ at $d \!=\! -1\,$pc.

%
\begin{deluxetable}{rcccc}[t!]
\tabletypesize{\scriptsize}
\tablecaption{\label{tab:pwn}}
\tablewidth{\columnwidth}
\tablehead{\colhead{$d$~[pc]} & \colhead{$B$~[mG]} & \colhead{$\mathcal{L}_{e^\pm}$~[$10^{35}\,$erg~s$^{-1}$]} & \colhead{$\alpha$} & \colhead{$E_c$~[TeV]}   }
\startdata
\hline \vspace{-0.2cm}\\
%
%
0.3	& 2		& 1.19	& -1.8	& 500 \\
-1	& 0.96	& 1.08	& -1.8	& 250 \\
-1	& 0.115	& 3.05	& -2.0	& 300 \\
\hline\vspace{-0.3cm}
\enddata
\tablecomments{Parameters for our TeV gamma-ray scenarios in Fig.~\ref{speca}.}
\end{deluxetable}
%

We begin at the center, $d \!=\! 0.3\,$pc.  The above equations give $B_1(0.3~{\rm pc}) \!=\! 3.2\,$mG and $B_2(0.3~{\rm pc}) \!=\! 1.28\,$mG.  We choose an intermediate value of 2~mG which well illustrates the general behavior.  The $e^\pm$ luminosity, $\mathcal{L}_{e^\pm}$, and other parameters for this and subsequent scenarios are collected in Table~\ref{tab:pwn}.  In Fig.~\ref{speca} we show the results of setting the $e^\pm$ injection to match the X-ray data.  Clearly, such a strong magnetic field has a much greater influence on the evolved $e^\pm$ spectrum than does the photon field at the energies of interest here.  As such, the gamma-ray flux comes in well below the TeV data.

We turn to scenarios assuming locations along the same line of sight at a distance of 1~pc from Sgr~A$^*$, both in front and to the rear of Sgr~A$^*$, with much weaker magnetic fields, $B_1(1~{\rm pc}) \!=\! 0.96\,$mG and $B_2(1~{\rm pc}) \!=\! 0.115\,$mG.  Taking these for both locations, the total energy loss rate is similar; however, the observed IC flux depends on the positioning of the source and scattering backgrounds relative to the observer.

In Fig.~\ref{speca} we show the spectra from the 1~pc model situated behind Sgr~A$^*$.  The gamma-ray flux using the larger field is again well below the TeV data.  Not shown in Fig.~\ref{speca} are the spectra calculated from the same distance only to the front of Sgr~A$^*$.  Using the same parameters, these are similar in shape with a gamma-ray flux normalization lower by a factor of $\sim\,$2.

The 0.115~mG case does yield a flux with the basic features, at the cost of introducing a field lower by a factor of nearly $\sim\,$100 than that implied by the GC magnetar.  Note that the X-ray flux here is not a power law, but curves through the {\it Chandra} flux due to our choice of exponential cutoff in combination with the other parameters.  As discussed by \citet{Hinton2007}, with the X-ray and gamma-ray fluxes both derived from the same $e^\pm$ spectrum, which is shaped by the combined action of both loss mechanisms, it is thus relatively harder due to the KN suppression of IC.

Fitting to the {\it Chandra} data then essentially decides the shape of the gamma-ray curves and leaves little flexibility in, e.g., either forcing a sharper gamma-ray break above 10~TeV or simultaneously accommodating a hard X-ray spectrum into the range now measured by {\it NuSTAR}.  This can seen via the characteristic $e^\pm$ energy $E_e$ emitting X-rays with $E_X$,
\begin{equation}
       E_e \sim 20\, \left(\frac{E_X}{10\,{\rm keV}}\right)^{1/2} \left(\frac{{\rm mG}}{B}\right)^{1/2} \, {\rm TeV}
    \,.
\label{Echar}
\end{equation}

While the last model displays a gamma-ray flux with desirable properties, the field strength runs into difficulties accounting for the spectral cooling evident in {\it Chandra} data.  Also, if we had chosen the same field strength to describe the Cannonball ($\sim\,$20~pc projected) this would have been well below the derived $>\,$0.3~mG field \citep{Nynka2013,Zhao2013}.  We would thus be in a situation where the synchrotron losses experienced by the PWN apparently more proximate to the GC need be roughly an order of magnitude smaller.

\section{Viability of a Galactic Center PWN}
\label{viab}
This one-zone description serves to narrow down the requirements for both the synchrotron and inverse Compton fluxes to be separately compatible with the X-ray and gamma-ray data.  However, lacking direct evidence via pulsed emission, a primary question remains as to whether G359 is really a PWN.  If so, can the X-rays and gamma rays be consistently explained?  Our aim is not to model G359 in great detail, but to consider a few basic arguments in examining its viability as a PWN and a TeV gamma-ray source.

First, we recognize that the thermal and magnetic pressures are rather large in the central parsec and might rival the ram pressure for even a relatively large pulsar velocity.  Assuming density $n \!=\! 26 (0.4\,{\rm pc}/r)\,$cm$^{-3}$ and $T \!\sim\! 10^7\,$K \citep{Baganoff2003,Muno2004,Wang2013}, we estimate the thermal and magnetic pressures
\begin{align}
   P_T \!=\! 2 \, n\, k_B T &\!\sim\! 10^{-7}\,{\rm erg~cm}^{-3}\,, {\rm at~0.3~pc} \notag \\
         &\!\sim\! 3 \!\times\! 10^{-8}\,{\rm erg~cm}^{-3}\,, {\rm at~1~pc}; \notag \\
   P_B \!=\! B^2/8 \pi &\!\sim\! 4 \!\times\! 10^{-8}\,(B/1~\rm{mG})^2\,{\rm erg~cm}^{-3}. \notag
\end{align}
For a pulsar speed $v_p$, we estimate the ram pressure as
\begin{align}
   P_R \!=\! n \, v_p^2 &\!\sim\! 1.4 \!\times\! 10^{-7}(v_p/500\,{\rm km~s}^{-1})^2\,{\rm erg~cm}^{-3}, {\rm at~0.3~pc} \notag \\
         &\!\sim\! 4.3 \!\times\! 10^{-8}(v_p/500\,{\rm km~s}^{-1})^2 {\rm erg~cm}^{-3}\,, {\rm at~1~pc}. \notag
\end{align}
These pressures could then all be comparable if the G359 morphology implies $v_p$ greatly exceeding the $10^7\,$K sound speed, $c_s \!\sim\! 380\,$km~s$^{-1}$.

Lacking timing information about the pulsar, we estimate its power, $\dot{E}$, starting from the particle populations injected to account for the X-ray data (recorded in Table~\ref{tab:pwn}), which require $\mathcal{L}_{e^\pm} \!\gtrsim\! 10^{35}\,$erg~s$^{-1}$.  If we assume that this results from an acceleration efficiency $f_p$ at a shock, then
\begin{equation}
      \dot{E} \!\gtrsim\! 10^{35}/f_p\,{\rm erg~s}^{-1}\,.
\label{edot}
\end{equation}
The standoff distance of the termination shock of the PWN can be estimated by balancing the wind power with external pressure (e.g., \citealt{Rees1974,Gaensler2006}) as
\begin{align}
      r_s &\!=\! (\dot{E}/4 \pi c \, P)^{1/2} \\
            &\!\sim\! 5 \!\times\! 10^{15} (\dot{E}/10^{36}\, {\rm erg~s}^{-1})^{1/2} (10^{-7}\,{\rm erg~cm}^{-3}/P)^{1/2}\,{\rm cm} \notag
\end{align}
or $\sim\!1.7 \!\times\! 10^{-3}\,$pc.  This is $\sim\,$0.04 arcsec, consistent with no obvious extended head structure being visible in X-rays.

Since $v_p$ is likely not much greater than $c_s$, we assume the distance to the back of the termination shock is comparable to $r_s$, $r_b \!\sim\! r_s$.  The field just beyond $r_s$ can be estimated at small wind magnetization (ratio of magnetic to particle energy fluxes), $\sigma$, following \citet{Kennel1984} as $B_s \!\approx\! 3 (\sigma  \dot{E}/r_s^2 c)^{1/2}$ or $\sim\! 2.4\,\sigma^{1/2}$mG.  For $\sigma \!=\! 0.01$, the 1D MHD modeling of \citet{Kennel1984} displays a field strength that subsequently rises by $\sim\! 2$ up to $\sim\! 4\, r_s$ before slowly declining as $\sim\! r^{-1}$.  However, the outflow confinement should arrest the decline relative to this isotropic limit.

The axisymmetric MHD simulations including ram pressure of \citet{Bucciantini2005} show a PWN structure separated into a channel with bulk velocity $v_1 \!\approx\! 0.1 \!-\! 0.5\,c$ surrounded by a flow with $v_2 \!\approx\! 0.8 \!-\! 0.9\,c$.  In these models, the PWN tails are roughly cylindrical with radius $R_{\rm tail} \!\sim\! 4 r_s$ and pressure $P_{\rm tail} \!\sim\! 0.02 \, n \, v_p^2$.  This would be smaller than $P_T$ (and near Sgr~A$^*$ even $P_B$) for the nominal parameters above.  This could be used to argue that G359 is well beyond the GC.  However, if this pulsar is the only one satisfying the criteria for observation as a PWN in the central parsec, it may well possess a larger power and above average velocity.  If traveling into a headwind, which for gas orbiting the GC could be substantial, it would be easier to fit into this picture.

\begin{figure*}[t!]
\vspace*{-0.2cm}
\hspace*{0.7cm}
\includegraphics[width=1.87\columnwidth,clip=true]{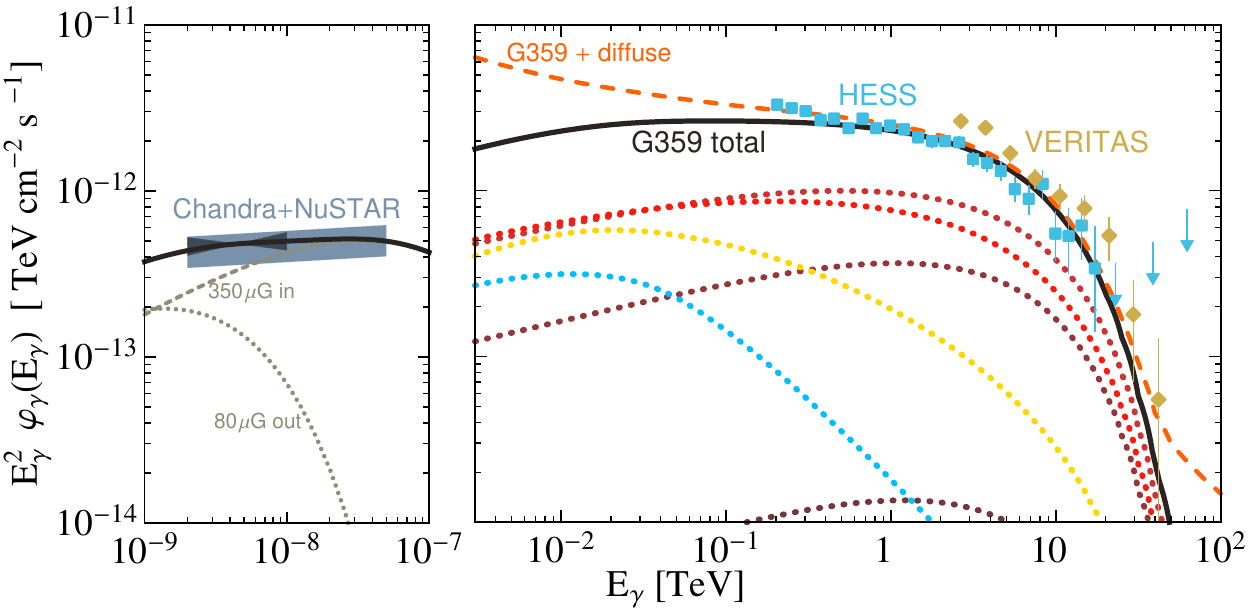}
\caption{Emission spectra for the two component G359 model.
{\it Left:} Synchrotron flux from the inner 350~$\mu$G ({\it dashed line}) and outer 80~$\mu$G ({\it dotted}) regions.
{\it Right:} Inverse Compton from a distance of 1~pc from Sgr~A$^*$ (to the rear) broken down into individual scattering background components used in Fig.~\ref{bEfig} and their total ({\it solid line}) that includes $\gamma\!-\!\gamma$ attenuation, along with the addition of a diffuse component ({\it dashed line}; see \S\ref{viabT}).
\label{spec1}}
\end{figure*}

The exact mechanism(s) of particle acceleration in PWNe remains undetermined, although recent advances offer insight into conditions that may be favorable.  If the magnetic energy content of the wind is much lower than that in particles, due to dissipation prior to $r_s$, Fermi acceleration can redistribute energy amongst the particle population, with total number and energy being fixed.  Simulations of relativistic pair shocks indicate that this is not very effective at producing hard particle spectra \citep{Sironi2013}.  Alternatively, a high magnetization could permit efficient magnetic reconnection (e.g., \citealt{Sironi2014}; \citealt{Guo2014}; \citealt{Werner2014}), transfer of field energy to particles, and a hard final spectrum for the population (total particle number still fixed with a increased mean energy).  In either case, the where and how of particle energization from the field determine the outcome.

We proceed from $\dot{E}$, which implies a \citet{Goldreich1969} potential $\Phi_{\rm GJ}$ and maximum energy
\begin{equation}
      E_{\rm GJ}^{\rm max}  \!=\! e \Phi_{\rm GJ} \!=\! e\,(\dot{E}/c)^{1/2}\,.
\label{edot1}
\end{equation}
This also implies a particle flux $\dot{N}_{\rm GJ} \!=\! c\, \Phi_{\rm GJ}/e  \!=\! (c \dot{E}/e)^{1/2}$ that we relate to the total wind flux via
\begin{align}
      \int dE\, \frac{dN_{e^\pm}}{dE\,dt}& = \dot{N} = \mathcal{M} \dot{N}_{\rm GJ}\,, \\
      \int dE\,E \frac{dN_{e^\pm}}{dE\,dt} &= f_p \dot{E}\,,
\end{align}
with pair multiplicity $\mathcal{M}$ and a bulk $\Gamma_{\rm wind} \!\sim\! \dot{E}/(m_e c^2 \dot{N})$.

As determined above, $\alpha \!\sim\! -2$ is a starting point for the injected spectrum.  This could plausibly be reached in acceleration via forced reconnection, which requires $\sigma \!\gtrsim\! 10$ at $r_s$ \citep{Sironi2014}.  Traditionally, it has been a problem to dissipate the field content of the wind to get $\sigma \!\ll\! 1$ to match the detailed observations of the Crab nebula (\citealt{Kennel1984}; cf., \citealt{Porth2013}).  This could arise if a striped wind (with fields alternating polarity with radial distance) is able to merge and reconnect prior to $r_s$ (e.g., \citealt{Coroniti1990,Kirk2003}).  For a sufficient travel time in the wind frame to merge, $\Gamma_{\rm wind} \!\lesssim\! [(r_s/\pi r_{\rm LC})(v_w/c)]^{1/2}$ \citep{Arons2012}, for relative velocity $v_w$, light cylinder radius $r_{\rm LC} \!=\! c/\Omega_p \!\simeq\! 1576\, (P_p/33~{\rm ms})$~km, and pulsar period $P_p$.

A large pressure and an $\dot{E}$ near the minimum required by the $e^\pm$ result in a small $r_s$, while $P_p \!\approx\! 200\,$ms would place G359 in the cloud of $\dot{E} \!\gtrsim\! 10^{35}\,$erg~s$^{-1}$ {\it Fermi} pulsars \citep{Abdo2013}, so $\Gamma_{\rm wind} \!\lesssim\!  10^3 (v_w/c)^{1/2}$ for merging.  This appears to be at odds with the need for TeV $e^\pm$ and $\Gamma_{\rm wind} \!\sim\! 3 \!\times\! 10^4$ for $\dot{E} \!=\! 10^{36}\, {\rm erg~s}^{-1}$ even with $\mathcal{M} \!=\! 10^5$, so $\sigma$ could remain large until $\sim\! r_s$.  \citet{Sironi2011} present a criterion of $4 \pi \mathcal{M}\, r_{\rm LC}/r_s \!\gtrsim\! 10$ for efficient acceleration via reconnection, though they note that this requirement is generally rather difficult to accommodate in PWNe.  While it is not clear how important this is to meet, this value can be approached pushing to $P_p \!\approx\! 300\,$ms, $r_s \!\lesssim\! 0.5 \!\times\! 10^{15} \,$cm, and $\mathcal{M} \!\gtrsim\! 10^5$.

\section{Emission from a PWN at the Galactic Center: II}
\label{viabT}
In light of the above, we consider a modified, yet still fairly simple, scenario for G359.  In this, we assume that the accelerated $e^\pm$ enter a region with a 350~$\mu$G field in which most synchrotron losses occur, with a $\sim\! 1\,$lyr length comparable to the extent of G359.  Afterwards, they enter an 80~$\mu$G field, owing to either a cessation of a coherent nebula due to the high external pressure or simply reflecting a declining field strength in an extension of the nebula.

Fig.~\ref{spec1} shows the results of using $\alpha \!=\! -1.9$, $E_c \!=\! 100\,$TeV (well below $E_{\rm GJ}^{\rm max}$ for the nominal $\dot{E}$), and injecting $\mathcal{L}_{e^\pm} \!=\! 3.5 \!\times\! 10^{35}\,$erg~s$^{-1}$.  We again assume $\tau \!=\! 10^3\,$yr, which is less relevant for X-ray synchrotron due to the short cooling time at the energies producing X-rays, although we cut the 80~$\mu$G X-ray fluxes at 30~yr to limit comparison to roughly the same spatial region.

Due to the larger synchrotron losses in the high-field region, the high-energy portion of the spectrum is burnt off prior to entering the lower field where the relevance of IC increases.  The resulting gamma-ray spectrum then displays a stronger than exponential cutoff, which could account for the puzzling steep drop in the 10~TeV data.  The solid line in Fig.~\ref{spec1} includes $\gamma\!-\!\gamma$ attenuation from Fig.~3 of \citet{Kistler2015}, while the dashed line adds a flux from hadronic interactions of an $E^{-2.5}$ proton spectrum (using the model of \citealt{Kelner2006} as in \citealt{Beacom2007}) as suggested by \citet{Viana2013} if part of the diffuse TeV component identified by HESS in \citet{Aharonian2006} contributes to the source flux.

While the low energy extent of this $e^\pm$ spectrum does not affect the total energetics much, it does greatly change the implied $\mathcal{M}$.  Allowing the $\alpha \!=\! -1.9$ spectrum to continue down to $E_e \!=\! 1$~GeV leads to $\mathcal{M} \!\sim\! 3 \!\times\! 10^4$, close to the value from above.  \citet{Wang2006} and \citet{Hinton2007} suggested that a lack of radio emission at G359 implies a cutoff at the low energy end of $\gtrsim \! 5 \!-\! 50\,$GeV.  This is really all that is required to account for the X-ray and TeV emission, though leading to a much lower $\mathcal{M} \!\sim\! 100$.  If radio is a problem at large downstream distances, the PWN otherwise needs to be young to avoid an accumulation of radio emitting $e^\pm$.

In contrast to previous assumptions, we conclude that in this scenario the TeV gamma-ray emission should {\it not} trace the X-ray emission of G359, since these are predominantly occurring in two distinct regions.  Since the gamma rays arise much farther down the flow, they would be separated from both the PWN position and Sgr~A$^*$.  The exact spatial structure in the low-field region is left to the vagaries of the local magnetic field in which the electrons are presumed to then propagate, which is discussed in relation to other data in \citet{Kistler2015}.

Although we have not explicitly constructed a dynamic model, there should be a diminution of the X-ray spectrum along with a softening, consistent with the steepening seen by \citet{Wang2006}.  Granted, we could also be assuming too much in interpreting from a single $e^\pm$ spectral index, as in principle the total spectrum can be the superposition of two processes, e.g., $E^{-1.5}$ and $E^{-2.5}$ over the energy range of interest, perhaps due to latitudinal dependence of the wind properties.  Also, a flow separated into two channels with distinct velocities, as seen in \citet{Bucciantini2005}, allows additional freedom to match details of the data.

From the $e^\pm$ luminosity bound on $\dot{E}$, we also obtain a first estimate of the pulsed gamma-ray emission of G359.  This generally depends on the pulsar magnetic field geometry and the sightline to the observer.  The lack of radio may imply a radio-quiet pulsar, and these have been detected by {\it Fermi} up to $\dot{E} \!\sim\! 5 \!\times\! 10^{36}\,$erg~s$^{-1}$ \citep{Abdo2013}.  We show in Fig.~\ref{speca} the isotropic GeV flux for $\dot{E} \!\sim\! 3 \!-\! 30 \!\times\! 10^{35}\,$erg~s$^{-1}$ using the estimated average spectra for this $\dot{E}$ range from \citet{OLeary2015,OLeary2016}.  This can be somewhat higher if beaming is relevant and the beam observable (e.g., \citealt{Watters2009,Pierbattista2015}), although the pulsed flux is likely well below the {\it Fermi} GC source 3FGL J1745.6--2859c \citep{Acero2015}, leaving room for other contributions, such as less powerful, but more numerous, pulsars located in this region.

\section{Discussion and conclusions}
\label{concl}
Besides novel features imposed by the properties of its surroundings at the Galactic Center, such as a potentially large external magnetic field, the putative pulsar wind nebula G359.95--0.04 is of great interest due to being situated within a TeV source also coincident with Sgr~A$^*$.  A lack of strong direct evidence for Sgr~A$^*$ makes it paramount to understand G359 in case the supermassive black hole actually is the accelerator yielding the gamma rays.

The several mG ambient magnetic field strength implied by data from the nearby Galactic Center magnetar, SGR J1745-29, is rather high (and possibly stronger if field reversals mask an even larger value; \citealt{Eatough2013}).  The cooling times for high-energy electrons in such fields are quite short and there would be a meager TeV gamma-ray yield despite the GC photon field.  Thus, if G359 were as close to Sgr~A$^*$ as its projected 0.3~pc, one would be left with needing a much weaker field within the PWN, which would be rather peculiar in comparison to objects in the outer galaxy, where it is easy to assume that the field resulting in the synchrotron radiation is at least as large as the ambient ISM field.

If one only wished to account for the G359 {\it Chandra} and {\it NuSTAR} data, the simplest explanation may be to assume that the object is significantly beyond the Galactic Center and only apparently near Sgr~A$^*$ by coincidence, thus avoiding contending with the harsh environment.  Moving the PWN further out, the ambient field may be weaker; however, the field must still be strong enough to explain the observed X-ray spectral variation, and we can use this emission to gauge the $e^\pm$ flux and spectrum.  Doing so, and assuming a uniform field and photon background similarly to \citet{Wang2006} and \citet{Hinton2007}, the gamma-ray flux would still come in well below the TeV data since there would no longer be a large photon density to result in a large $L_{\rm TeV}/L_{\rm keV}$ ratio.

Making the added assumption of a spatially varying field, we found that rough agreement can be obtained in the context of our model for a relatively narrow range of parameters, including distance, luminosity, spectra, field strengths, and assumptions about the PWN structure.  For instance, in the model displayed in Fig.~\ref{spec1}, the $e^\pm$ emit most of their TeV gamma rays from within a 0.08~mG field, roughly two orders of magnitude smaller than the central magnetar value.  If we take the field strength of $\gtrsim\! 0.05\,$mG derived by \citet{Crocker2010} as pervading the central $\sim\,$100~pc as a lower bound, there would not be much additional leeway in moving the PWN much further than $\sim\! 1\,$pc due to the declining background photon densities.

Considering as well the interest in the GC for reasons ranging from dark matter annihilations yielding GeV gamma rays (e.g., \citealt{Abazajian2014,Daylan2014,Calore2014,Ajello2016}) and the origin of the {\it Fermi} bubbles \citep{Su2010,Ackermann2014}, as well as understanding multi-TeV $e^\pm$ production by pulsars throughout the galaxy (see, e.g., \citealt{Yuksel2009,Kistler2012}), it is important to understand the $e^\pm$ population present.  It would be of great benefit to be able to perform a detailed spatial/spectral analysis as for other objects (e.g., \citealt{Van2011,An2014,Gelfand2015}) or to examine X-ray data for time variability as a sign of bulk motion.

Also, a variety of models yield TeV gamma rays either from Sgr~A$^*$ or via cosmic ray interactions in the immediate environment (e.g., \citealt{Atoyan2004,Aharonian2005,Quataert2005,Liu2006,Ballantyne2007,Ballantyne2011,Linden2012,Kusunose2012,YusefZadeh2013,Fujita2015,Kistler2015c}).  Many of these yield neutrinos \citep{Crocker2005,Kistler2006}, which do not result from a leptonic PWN model.  Searches using IceCube  \citep{Aartsen2013,Aartsen2013b,Aartsen2014} or a km$^3$ Mediterranean detector \citep{Coniglione:2015aqa} may then be decisive.  This itself would imply a TeV connection to Sgr~A$^*$ and proton acceleration by supermassive black holes, providing a crucial link between cosmic rays and neutrinos \citep{Kistler2014,Kistler2016}.

How might we proceed directly on G359?  Radio observations have revealed the proper motion of the GC magnetar relative to Sgr~A$^*$ to be $\sim\!6\,$mas~yr$^{-1}$, a $\sim\!236\,$km~s$^{-1}$ transverse velocity \citep{Bower2015}, while pulsar speeds of $v_p \!\gtrsim\! 1000\,{\rm km~s}^{-1}$ are not unprecedented (e.g., \citealt{Chatterjee2005,Ng2012,Halpern2014}).  For G359, $v_p\,$sin$\,i \!\sim\! 700\,{\rm km~s}^{-1}$ would imply $\sim\!0.2\,''$ per decade, measurement of which would be strong evidence of a fast pulsar, perhaps even in X-rays over a decent interval.\\

%
We thank Jon Arons, John Beacom, Roger Romani, Carl Farbeman, and Hasan Yuksel for useful discussions and advice.
MDK acknowledges support provided by Department of Energy contract DE-AC02-76SF00515, and the KIPAC Kavli Fellowship made possible by The Kavli Foundation.


\end{document}